**Field Emission in CEBAF's Superconducting RF Cavities and Implications for Future Accelerators**

Jay Benesch
Thomas Jefferson National Accelerator Facility (Jefferson Lab)

**Abstract**

Field emission is one of the key issues in superconducting RF for particle accelerators.[1] When present, it limits operating gradient directly or via induced heat load at 2K. In order to minimize particulate contamination of and thus field emission in the CEBAF SRF cavities during assembly, a cold ceramic RF window was placed very close to the accelerating cavity proper. As an unintended consequence of this, the window is charged by field-emitted electrons, making it possible to monitor and model field emission in the CEBAF cavities since in-tunnel operation began. From January 30, 1995, through February 10, 2003, there were 64 instances of spontaneous onset or change in cavity field emission with a drop in usable gradient averaging 1.4 ($\sigma$ 0.8) MV/m at each event. Fractional loss averaged 0.18 ($\sigma$ 0.12) of pre-event gradient. This event count corresponds to 2.4 events per century per cavity, or 8 per year in CEBAF. It is hypothesized that changes in field emission are due to adsorbed gas accumulation. The possible implications of this and other observations for the International Linear Collider (ILC) and other future accelerators will be discussed.

**Monitoring Field Emission in CEBAF**

The CEBAF 5-cell cavity pair and helium vessel are shown schematically in figure 1. The features of interest for this work are the high resistance ($> 10^{12}$ ohms/square) cold ceramic RF window 7.62 cm from the beam axis, a fundamental power coupler (FPC) with significant magnetic dipole field, and sensors attached to the waveguide at room temperature. The FPC induces a transverse kick of ~20 milliradians–MeV/c when the electron is on-crest in the adjacent accelerating cell, 147º RF phase away, and the cavity gradient as a whole is set at 7 MV/m.[2] While only modest trajectory modeling has been done[3], it is clear conceptually and has been demonstrated in vertical test dewar experiments that field emitted electrons from either cavity in a pair can reach and accumulate on the cold ceramic window.[4,5] The same set of vertical dewar experiments demonstrated that the interposing of an elbow or dogleg waveguide between the fundamental power (input) coupler flange and the ceramic window dropped the electron current to the window by three orders of magnitude.

During CEBAF commissioning, arc discharges were seen at the cold ceramic windows via photomultipliers and vacuum sensors attached to the warm-to-cold transition waveguide. These were verified with spectroscopic observation in vertical dewar tests[6] to occur at the ceramic and may be either surface flashover or punch-through. The latter is demonstrated by leak testing – most of the cold ceramic window assemblies in the accelerator now have small holes. In vertical tests, with a picoammeter available to monitor field emission current to the window, the discharges occurred at roughly constant window charge.[5] There is no way to monitor field emission current directly in the accelerator as all the vacuum seals are metal and there is no access to the cold ceramic window. All that can be recorded is the occurrence of arc and vacuum faults and the gradient in the cavity at the time each fault occurred. Such records have been maintained since Jan. 30, 1995. The data set contains 427,400 RF faults through May 30, 2005.

My analysis assumes that the cold ceramic window is a perfect capacitor and that the charge at which a discharge occurs is constant. The interval between discharges is then inversely proportional to a constant field emission current. If the gradient setting is constant in the interval and RF is on throughout, one can easily apply a simple exponential or more rigorous Fowler-Nordheim[1] model to the data directly to obtain a field emission model for each cavity.

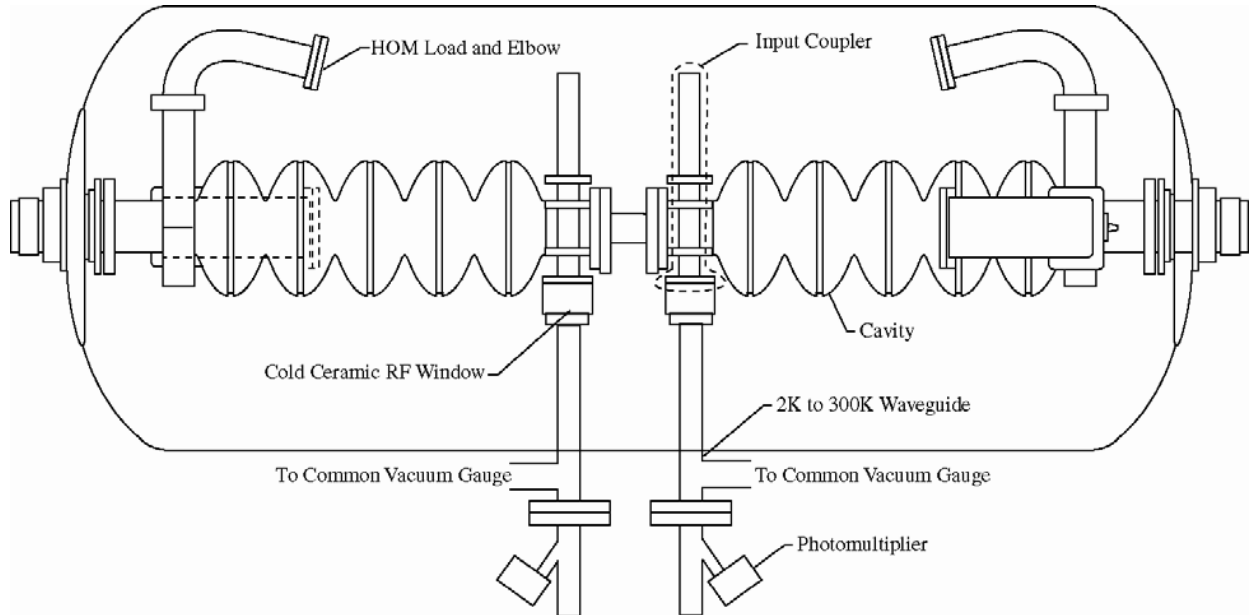

Figure 1. Cavity pair in helium vessel and room temperature sensors used in study.

The data is not perfectly clean so one pre-processing step and five data cuts are applied before statistical analysis. Until November 2004, there was no recordable signal giving RF-on time for each cavity. The pre-processing step approximates RF-on status by removing periods of 6+ hours in which no fault occurs anywhere in the machine from a running total of elapsed seconds. This assumes that all cavities are turned on and off at the same time, which is not the case – occasionally one linac is on and the other off for maintenance. This increases the noise in the data. The five cuts in the data and their justification are:

a. exclusion from analysis of faults with gradient under 3 MV/m due to limitations in RF control system stability which decrease fault interval for low gradients
b. exclusion from analysis of faults with intervals under 30 seconds due to variation in reset time from 7-30 seconds; reset was manual during most of the data collection
c. exclusion from analysis of faults with intervals more than 12 days (1036800 s) due to data plots suggesting that the assumption of perfect capacitors begins to break down at this interval. Data analyzed thus spans 4.5 orders of magnitude in interval.
d. exclusion from analysis of faults in which the gradient change from the preceding fault is more than 15%. There would be insufficient data to analyze if the assumption that the gradient is constant across the full interval were rigorously enforced. Both 10% and 15% cuts have been used with little difference in results. Since the gradient enters in the first power in the exponent in the simple exponential model and as the 5/2 power in the exponent in the Fowler-Nordheim model, no larger allowances were tested.
e. exclusion of simultaneous (within timing resolution) faults of cavities in multiple helium vessels as due to beam strike or control system effects rather than field emission.

Photomultiplier (PMT) and vacuum sensors were mentioned above. The first is termed the arc detector and is a simple threshold detector – if a PMT signal greater than a fixed level is detected for more than 0.5 ms, the RF is shut off and a fault bit set. The second is connected to a pair of cavities and the actual pressure archived as a function of time. About 20% of the accumulated RF faults show only vacuum faults and cannot be assigned to a single cavity, only to a pair.  Inclusion of these faults in the analysis of either member of the pair has always decreased correlation coefficients, so these faults are discarded.  Some fraction of these are likely accompanied by sub-threshold PMT signals and should be included but there is no way to determine which.  About 75% of the faults show simultaneous arc and vacuum faults.

About 5% of the faults show only an arc detector bit.  In early 2003 the archiving rate for the vacuum data was increased from 1 to 10 Hz.  This allowed the addition of a data pre-processing step which determines if there was a sub-threshold vacuum event at the same time as the arc detector fault and reclassifies about half of these 5% as true arcs.  An increase in vacuum reading at least equal to background is required for the reclassification.  When plotted, all such vacuum traces show classic burst and recovery patterns.  For the data set ending in February 2003, the author determined by visual inspection in the data exploration program JMP[7] which of the arc-detector-only faults would be included in the analysis.

The analysis which follows therefore begins with about 77% of the faults recorded and makes the cuts described above to this subset, ending with ~71% of the total faults.  Known noise sources include, as discussed above: imprecision in RF-on intervals, changes in gradient during intervals, variation in window charge at discharge, and nonassignable vacuum-only faults.

Figure 2 is a plot of the ln of the inverse fault intervals as a function of cavity gradient for cavity 0L031. This is the first cavity in CEBAF which varies in gradient; the two preceding cavities take the beam from 0.5 MeV to 5 MeV and are invariant.  Only seven points were removed by the data cuts discussed above.  Residuals of the fit are shown in figure 3 left.  The distribution is close to normal visually but does not satisfy the Shapiro-Wilk W test for normality[14].  Removing two outliers from the high side and eight from the low, followed by refitting, results in the residual distribution in figure 3 right, which is consistent with normality.

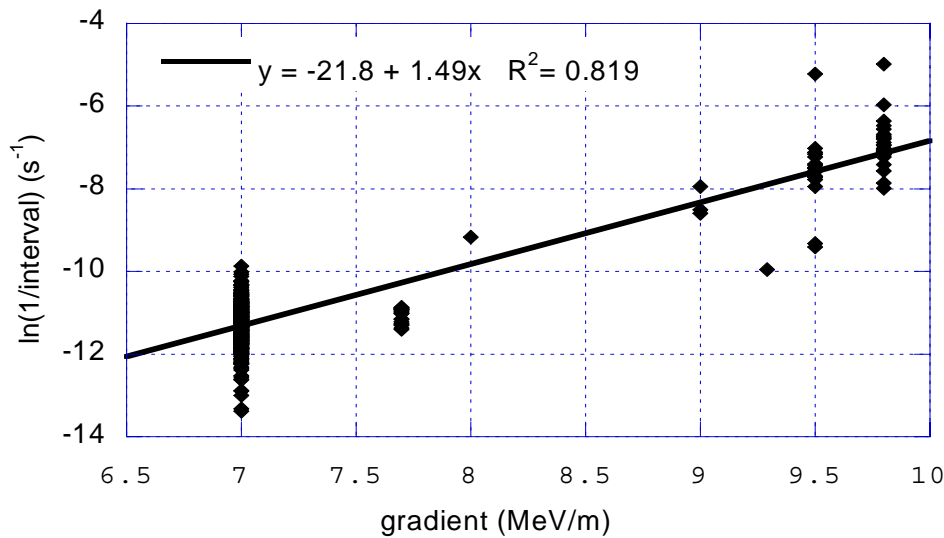

Figure 2.  Fit to 0L031 data, automatic cuts only

This labor-intensive process of exploration of the data sets for outliers and development of exponential and Fowler-Nordheim models for each of 338 CEBAF cavities has been repeated many times since the beginning of 1995.  The exponential models are used in a program which sets the gradient distribution along the linacs to minimize arc rate.[8]  The Fowler-Nordheim models were used through August 2003 during outlier removal as residuals tended to be closer to normal (Shapiro-Wilk W test).  After the hurricane-induced temperature cycle to room temperature in August 2003[9], such niceties were abandoned due to lack of time and only the exponential models were developed since those are the only ones used in machine setup.

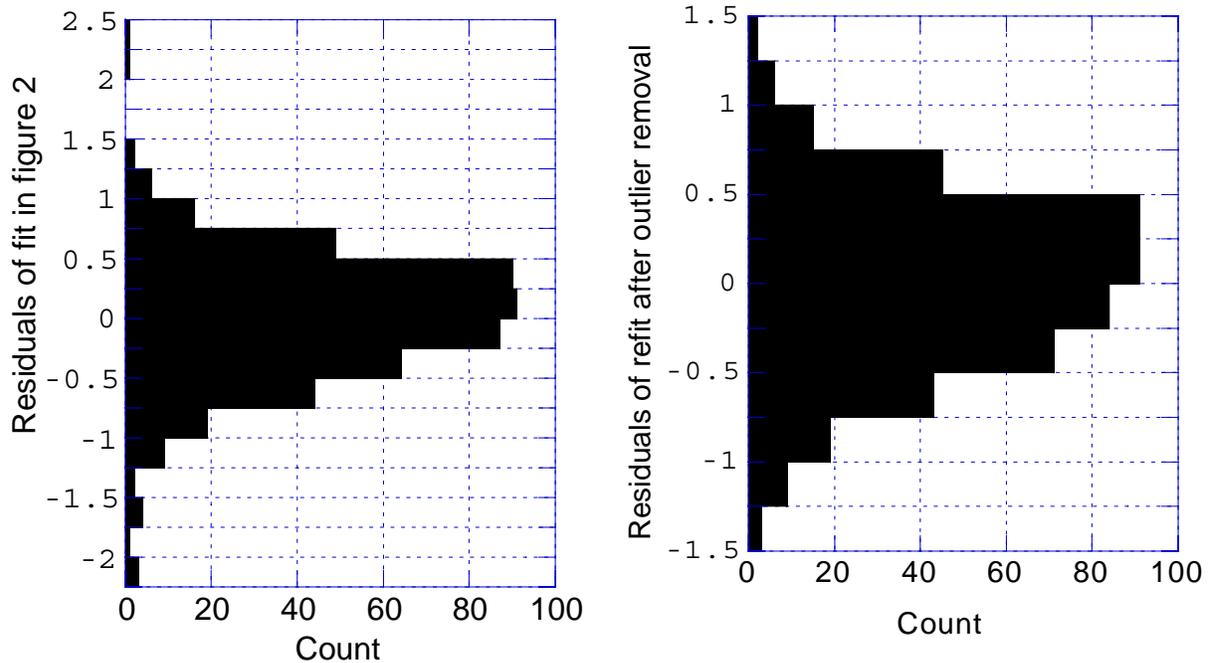

Figure 3. Residuals of fit in figure 2 before (left) and after (right) outlier removal.

**Abrupt Changes in Field Emission**

In figure 4 we show a change in cavity performance which does not correlate with any known external disturbance. Among the externally imposed changes which modify field emission and therefore mandate new statistical models are 30K warm-up, 300K warm-up and helium RF-discharge processing[10]. All of these will change adsorbed gas distribution, the lead hypothesis for the phenomenon discussed below. Inaccurately calibrated RF control hardware can also cause a change in apparent response as the gradient scale changes.

Sixty-four events with no external drivers were found with this distinct signature, including at least a factor of three change in fault interval at fixed gradient, in the data set encompassing Jan. 30, 1995, through February 10, 2003, all but one degradations. Most of the changes in interval were at least an order of magnitude. Some were associated with a period of RF-off for magnet or other maintenance in the tunnel, but no recorded maintenance of any sort on the cavity or RF system in question. Figure 5 is one such change. This count of 64 is the minimum for the period and corresponds to 2.4 per cavity-century, or eight per year in CEBAF with 338 cavities. For an ILC with 20,000 cavities, the toll might be 500/year if the phenomenon occurs. Without similar long term performance data from TTF or other high performing cavities, it is not known whether the ~ 2.5 MV/m loss at fixed fault interval seen in figure 4 is best viewed as an offset or fractional loss of previous behavior. The former is much preferred for ILC.

The cavity performance changes are not subtle, providing confidence that the 64 events selected are the minimum set. Maximum set would include perhaps half again this number. In figure 6 the distributions of absolute and fractional gradient loss for the 64 events are shown. Mean gradient loss is 1.4 ($\sigma$ 0.8) MV/m and mean fractional loss 0.18 ($\sigma$ 0.11). It is not clear whether the improvement in one cavity is real or a result of measurement error induced by unrecorded maintenance. Its inclusion renders the left distribution more consistent with normality. If excluded, losses are 1.4 (σ 0.7) MV/m and 0.18 (σ 0.10) fractional for the 63-event sets.

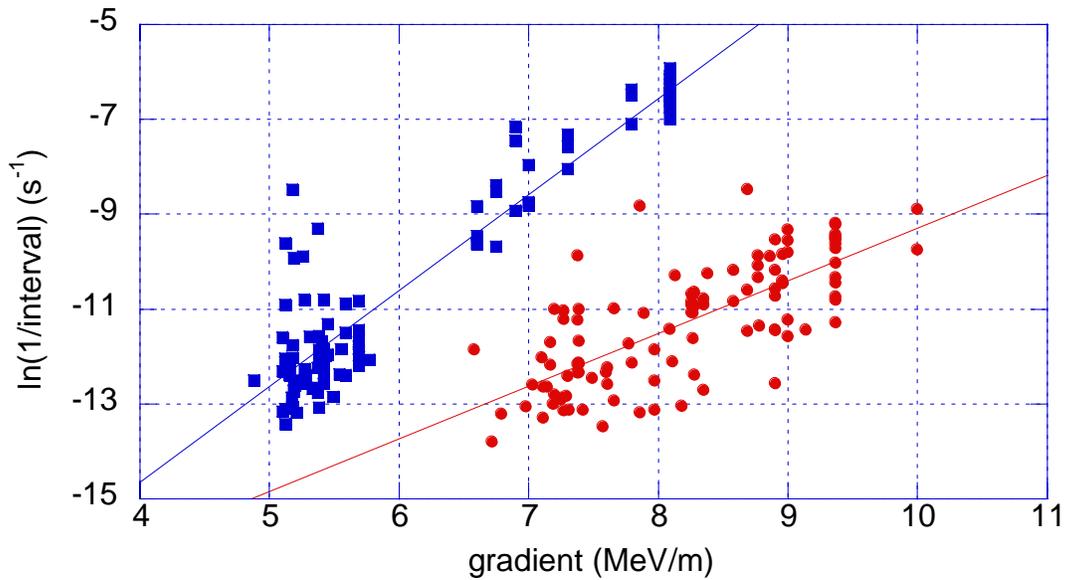

Figure 4. Blue squares are after 0440 9/21/2004; red circles before. Interval at 8.1 MV/m changed from ~80,000 seconds to ~500 seconds. Linear fits for the data sets before and after are shown. Cavity 2L145

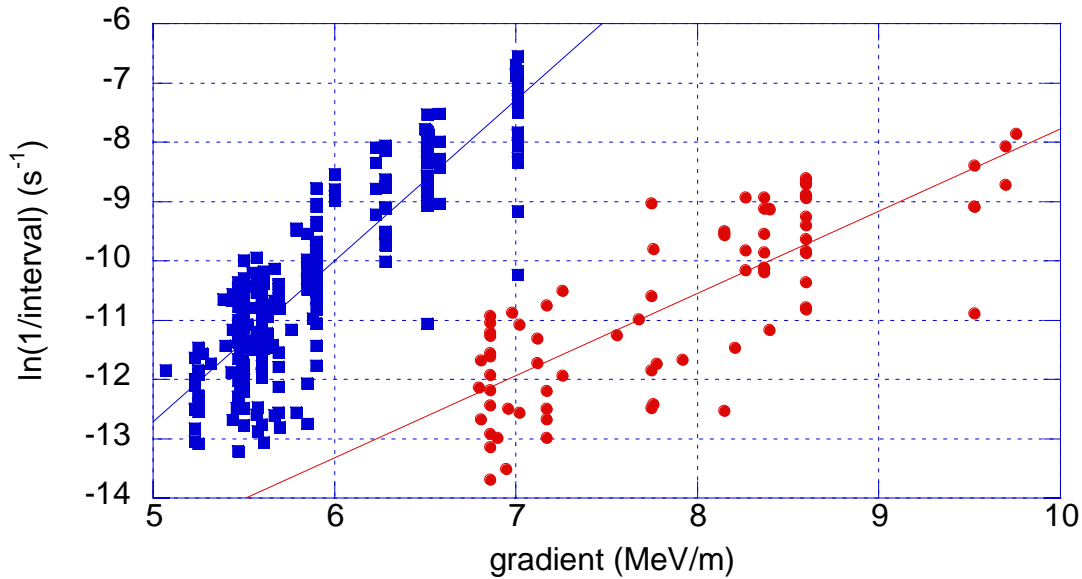

Figure 5. Data (blue squares) sloping across left of figure is after 5/18/2004; red circles before. This change in behavior occurred across a maintenance period during which nothing was done to the cavity except turn RF off and on. Such episodes are counted in deriving the 2.4/cavity-century value. Cavity 1L087

Seventeen such events were counted between the October 2003 resumption of operations after warm-up to ambient due to a hurricane[9] and the end of March 2005, a period of 18 months. Since fault data was not recorded for the first two years after original cooldown in CEBAF, it is not known whether this rate, almost 12/year vs 8/year before the hurricane, is appropriate for a new accelerator. The recent events cannot be quantified in the same manner as those in figure 6 because there isn't enough data to create models for most of the "before" states. For instance, there may be a dozen points at fixed gradient with an interval of 50000 seconds which suddenly changes to 100 seconds. An event has clearly occurred. "After" data at different gradients is available but only one gradient "before".

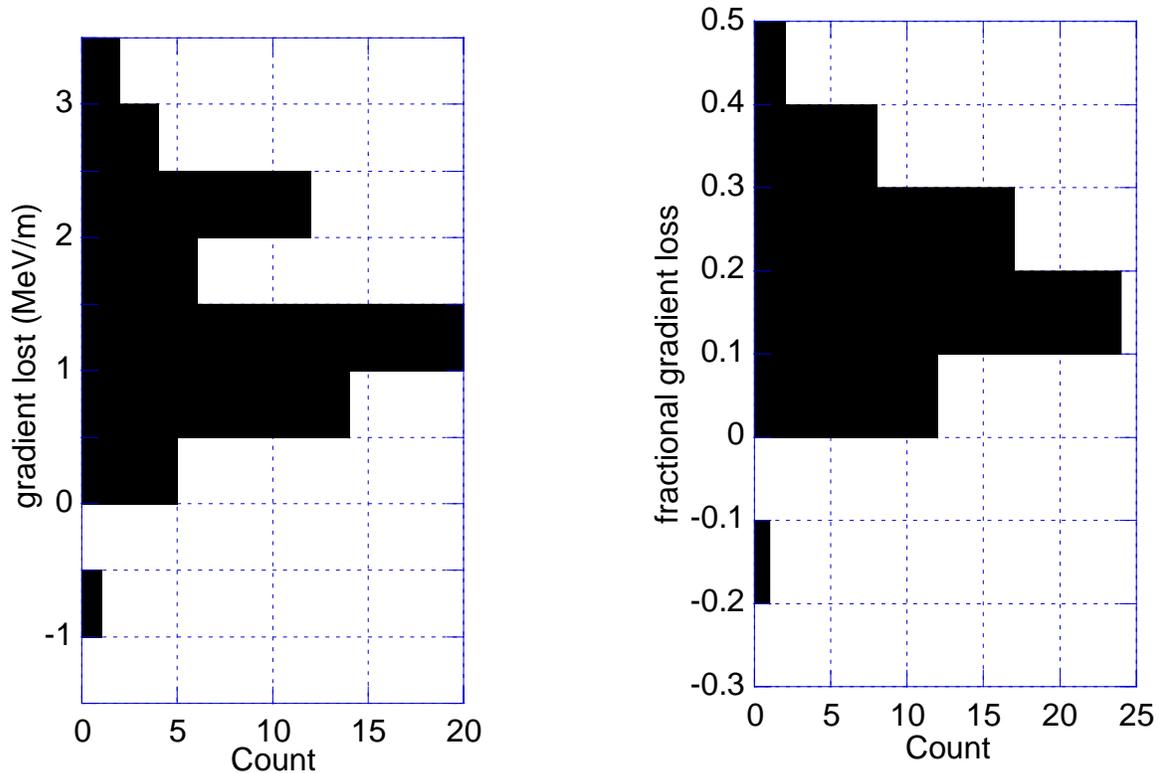

Figure 6. Abrupt loss distributions in two formats for January 30, 1995, through February 2003.

**Another phenomenon of interest: "fratricide"**

In addition to field emission in a cavity charging its own ceramic window and causing arcs, it is possible for an adjacent cavity to do so as well. Such "fratricide" may extend beyond the other member of a cavity pair to adjacent helium vessels. Fratricide is found and quantified by stepwise regression of multiple cavity gradients in JMP[7]. The check for fratricide is one of the reasons for the large amount of human input to the analysis. A tenth show such influence.

The author was convinced of the phenomenon's existence by the abrupt changes in the performance of cavities 6 and 7 in zone NL04 when an accidental introduction of $N_2$ into cavity 8 forced its gradient to drop from 10 MV/m to 5 MV/m. Retrospective analysis of previously misunderstood data showed that when cavity 8 was below 7.5 MV/m, fault vs gradient behavior in cavities 6 and 7 was consistent with field emission models. Figure 7 shows the data for cavity 6. Cavity 7 faulted so infrequently after cavity 8 was turned down that insufficient data is available for analysis. The physical mechanism of interaction between cavities which are not in the same pair is unknown. When fratricide is statistically found in cavity response and the culprit is located, a maximum culprit gradient is estimated and tested in the machine. Thirty-two models now in use are modified by fratricide.

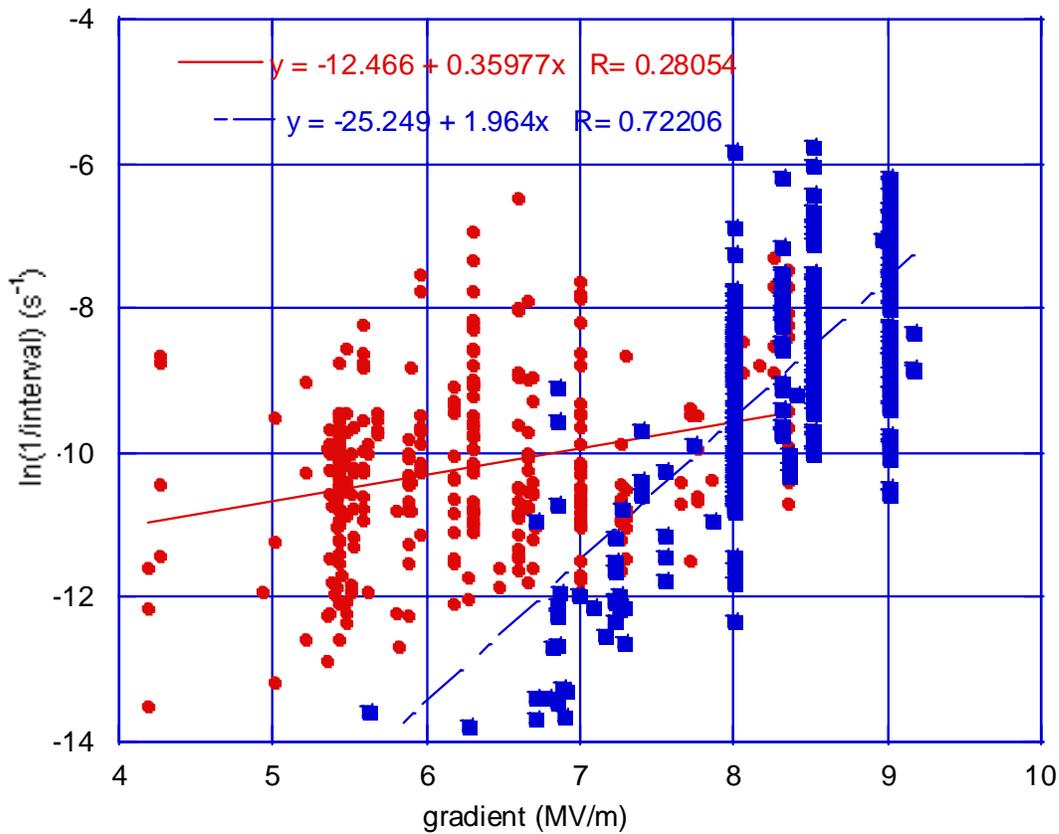

Figure 7 Fits of cavity 6 fault intervals with cavity 8 > 7.5 MV/m (red circles) and cavity 8 < 7.5 MV/m (blue squares)  The change in behavior is clear.

**Characteristics of the set of models now in use**

The ensemble of statistical models which were used in the operation of CEBAF[8] in early 2005 will now be characterized statistically.  Perhaps the most striking feature is an exceptional correlation of the slope and intercept of the exponential models (figure 8).  The physical source of this correlation is unknown.  It may be a function of cavity and FPC geometry.

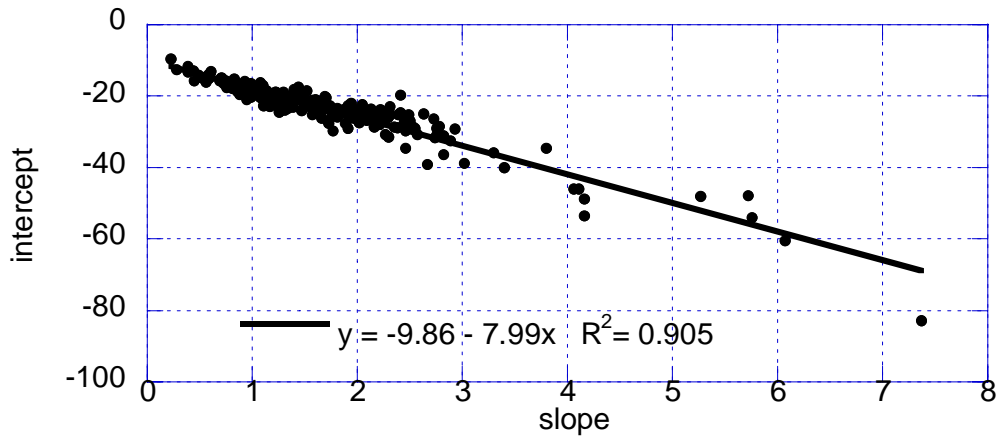

Figure 8 Intercept vs slope for 239 cavity models in use in CEBAF in June 2005

Statistical measures of the quality of these models are shown in figure 9. The t value is equal to parameter divided by standard error and so is a measure of the significance of the model. Minimum t value of 2, which for these samples corresponds to excluding the null hypothesis at P=0.05, is enforced for model use in the machine. 90% of the models used have t>7.9 for slope and t>10 for intercept: "7.9 sigma" and "10 sigma" in the usual physics parlance. $R^2$ is the square of the usual correlation coefficient. While some of these correlation coefficients are low, additional data can best be obtained if some model is used in setting up the gradient distribution in the accelerator. To increase data acquisition rate, up to half a standard error has been added to the slope values input to the code[8] used to establish the gradient distribution in the accelerator.

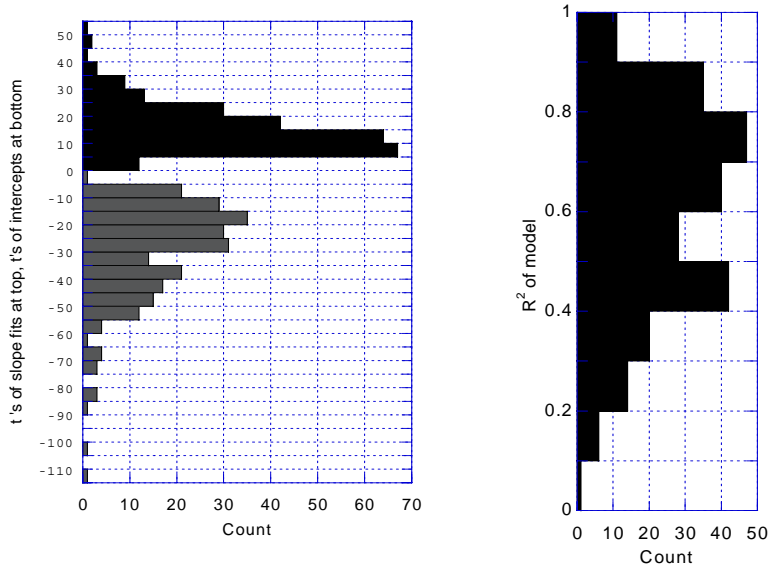

Figure 9. Statistical measures of model quality for 247 cavity models. t values for slopes and intercepts are at left and $R^2$ values at right. Statistical measures of model quality for 247 cavity models.

Models are lacking for the remaining 91 cavities in CEBAF for a variety of reasons: 8 failures requiring cryomodule remanufacturing; field emission onset above RF control or RF power upper bound; culprit in fratricide so upper bound set below effect of gradient on a cavity itself; room temperature RF window heating; insufficient data; and 7 injector cavities which haven't been pushed to limits due to fixed ratio between injector and linac energy.

**Attempts at automation of analysis**

The author has been working with a Jefferson Lab programmer to automate as much as possible of the analysis. All of the processing and data cuts described above are applied via a perl script. An additional cut proved necessary to deal with delays in resetting faults during short system problems: remove intervals under 1800 seconds with gradients below average. A list of cavities in which such points are more than 10% of the total is kept for manual check in JMP so abrupt changes as in figures 4 and 5 are not ignored. Four fitting routines from the open source statistics environment $R^{11}$ are then applied via an R script: standard least squares and three robust regression algorithms, M-estimator, MM-estimator and least trimmed squares. Numerical output is directed to a summary file and a graph of the data and the four fits is produced as a postscript file. If all four models agree and the statistical measures of the models indicate reliability, any of the models can simply be copied into the input file for the machine setup code. If they don't agree or if the statistical measures are poor, the plot is examined for signs of fratricide, sudden change of field emission characteristics, etc., and a decision made whether to repeat the analysis in JMP manually. This will reduce the effort needed in the future to develop new models when cryomodule perturbations force a change. The graph for 0L031 is shown in figure 10.

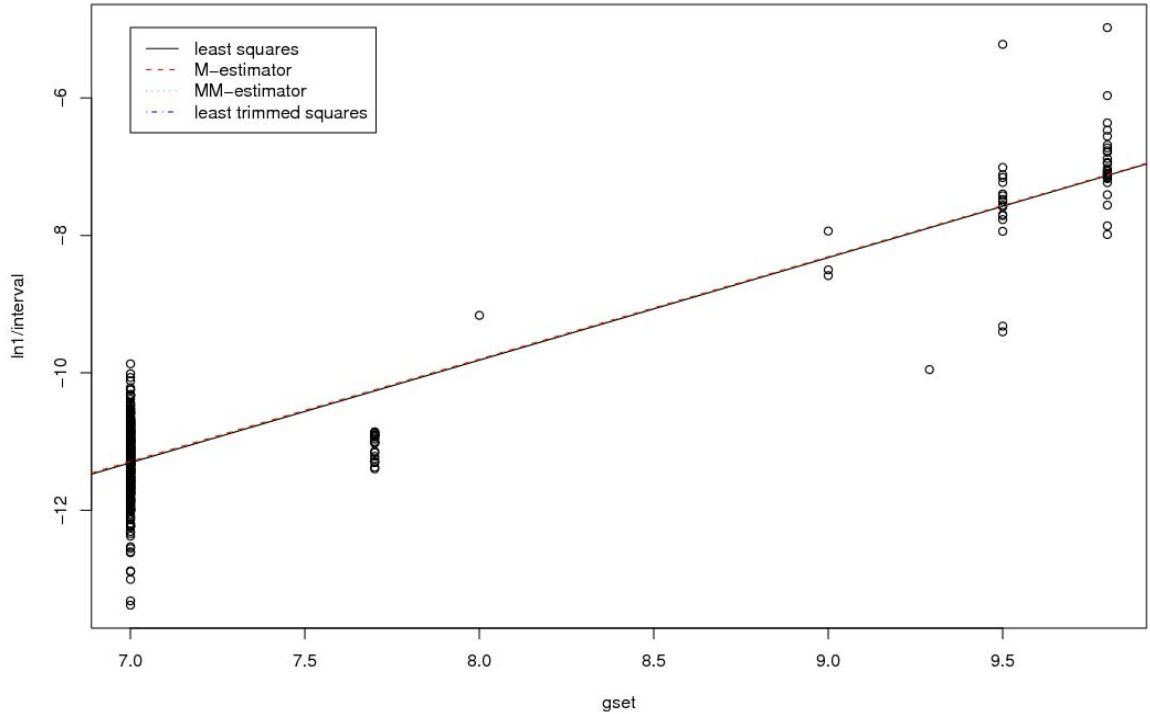

Figure 10. Cavity 0L031 data (compare figure 2) analyzed and plotted in R environment. Horizontal axis is the control system variable equivalent to gradient (MV/m). Compare with figure 2, from JMP.

**Effects of cryomodule perturbations**

Two recent events have resulted in major perturbations of multiple cryomodules. In August 2003, hurricane Isabel caused a four day power outage and all cryomodules warmed to room temperature.[9] In August 2004 maintenance of the main helium liquefier forced the cryo load to be shifted to a much smaller 4K refrigerator. The smaller capacity of this unit, in combination with main electrical substation maintenance, forced nine cryomodules to be warmed to room temperature. Change out of room temperature RF windows, requiring 30K cycle, occurred in four more modules in August 2004. Comparisons among models in March 2003, July 2004 and November 2004 were made. Due to poor statistics in the interval October 2003 - June 2004, comparisons with July 2004 models proved less than useful. In figure 11 gradients predicted to yield one day fault intervals are compared for March 2003 and November 2004 models in two ways, with intercept as a free parameter and with zero intercept forced. The second is not valid statistically but provides an estimate of the degradation due to the perturbation: 10%. The August 2004 excursions appear to have no effect on these comparisons as the same correlations are seen for the full set of cavities, the set which was unperturbed in August, and the set perturbed in August. Thus only the full set is shown.

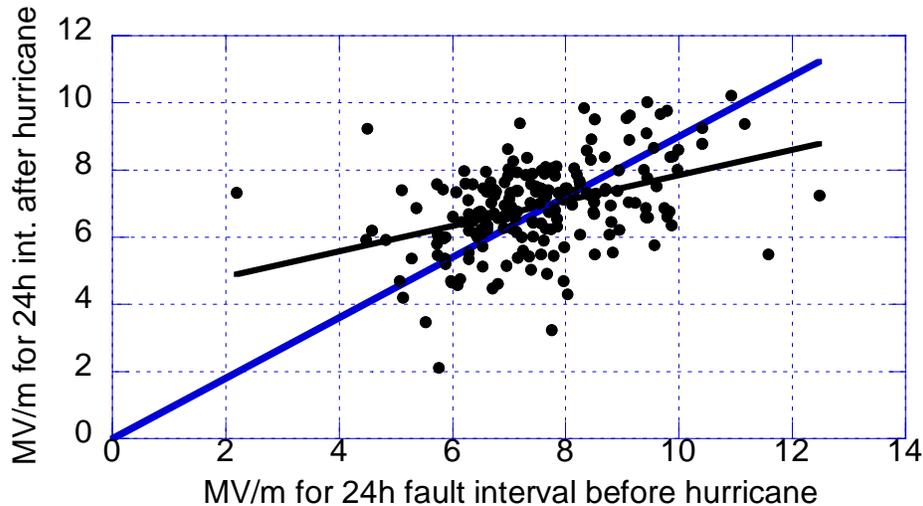

Figure 11. Comparison of pre- and post- hurricane gradients for one day fault interval by standard least squares (black) and fit forced through zero (blue). Latter has slope of 0.9. Former has equation y=0.38x + 4.06, R=0.42.

**Implications for Future Accelerators**

The more credible hypothesis for abrupt field emission changes advanced to date is an increase in geometrical electric field enhancement by "one more" gas molecule adsorbing to an asperity on the cavity surface. This mechanism should be independent of duty cycle, depending only on vacuum conditions. Leak rate of CEBAF cavity pairs was measured during superfluid helium vertical dewar tests via an integration method and average $4\times10^{-11}$ std cc/s [12]. The fact that CEBAF runs CW with RF on ~75% of the year and the ILC will run pulsed with ~1% duty cycle should be irrelevant for emission turn-on rate in this hypothesis. Particle migration is the other hypothesis remaining. Emitter turn-on has no correlation with cavity location within a cryomodule, but this result is not strong statistically with so few data. There are open gate valves with Viton gate seals at each end of each cavity pair and four ceramic HOM loads (figure 1) on each pair which might be sources of particles. The low correlation between pre and post hurricane gradients shown in figure 12 lends credence to the gas adsorbtion hypothesis because beam line vacuum remained in the molecular flow regime even at 300K, reducing the likelihood of particle motion. There's a significant difference here between CEBAF and future accelerators.

CEBAF runs with ~600W of 2K heating due to field emission at 5.8 GeV, or ~2.5W per cavity with field emission model. In vertical dewar tests, the author was able to run with up to 70W of field emission heating without quenching the cavity. Maximum field emission heat load allowable by the TESLA cryomodule design is not known to the author, but ~10W extra in one cavity probably won't choke the plumbing.

Improved surface preparation of the ILC cavities, including electropolishing, will reduce the asperity count per unit area orders of magnitude over that achieved in 1991–1993 in CEBAF. In the best case for ILC, this will cut the rate of abrupt changes to a level that is irrelevant for machine operation. If the 1995–2003 CEBAF rate holds in spite of this, undetected events could add up to 5 kW 2K heat load per year in ILC, but for a 1% duty cycle machine the increase in energy deposited to 2K is likely negligible. The heat load from new field emitters would be important for a future CW machine, for example ERL (energy recovered linac) based FELs or light sources, despite the fact that the number of cavities involved in any of these applications is small compared to the ILC.

The CEBAF event rate increased to 65 in 3.7 years after the cycle to ambient in September 2003, more than twice the rate cited above. The increase in rate is likely due to reduced pumpout time after the hurricane vs original construction, hence more gas present to move around and create emitters. Long term x-ray observation of TESLA modules at DESY and other labs would be useful but the number of cavities is too small to provide useful information before the planned completion of the ILC Conceptual Design.

During helium RF discharge processing to improve field emission[10] eight small G-M tubes were placed about the cryomodule and monitored. No pattern was found in x-ray emission with this small coverage. It was not possible to isolate the offending cavity to a single module using the x-ray detection alone, much less find the cavity. Observation of patterns of x-rays as a function of gradient was needed. The x-rays from some cavities didn't intercept the eight small detectors at all; their creation is inferred.

One mitigation implementation for this field emission change phenomenon in a future accelerator would require:
1. "4 $\pi$" x-ray detection in linacs
2. RF system capable of varying power to individual cavities
3. software to detect changes in x-ray patterns and use (2) to determine which cavity is at fault parasitically during normal operation. Energy lock assumed.

Item 1 can be retrofitted, for instance by applying sheets of plastic scintillator to the outside of the cryostat and guiding their light to appropriate detectors. Provision must be made in civil construction design for such a retrofit, e.g. cable and rack space. The decision whether to add (1) can be taken after observation of TESLA modules gains statistical significance. Item 2 must be part of the initial accelerator design. In-tunnel electronics, e.g. RF controls, would benefit from item 1 as they will have limits on integrated dose.[13]

Jefferson Lab is beginning a modest program to refurbish cryomodules. Improved surface preparation techniques will be applied with the goal of achieving at least 30 MV/m surface field (12 MV/m accelerating gradient) in refurbished cavities. The installed ensemble averaged 13 MV/m accelerating gradient with field emission in vertical test in 1991–1993. The best vertical test result was 21 MV/m, so 40 MV/m surface field average is not unrealistic after refurbishment with installed RF controls and klystons, roughly half that needed for ILC. Cavity surface field could approach ILC specs with improved RF systems.

Another part of this refurbishment is installation of a waveguide with dogleg between the cavity and the cold ceramic window. This has been shown to eliminate arc faults due to field emission in one cryomodule so equipped and so will eliminate the method for monitoring field emission used here. Radiation detectors can be placed as discussed above to monitor field emission in the absence of charge/discharge cycles on the cold ceramic windows. A more ambitious refurbishment program might apply two to four surface preparation techniques of varying complexity and cost to twenty to ten cryomodules each. Three years of observation would then produce statistically significant results on the efficacy, cost effectiveness, and longevity (against spontaneous field emission changes) of each treatment. Such information would be valuable for the ILC and other future accelerators.

**Summary**

Insights gained from a decade of monitoring and modeling field emission in CEBAF have been discussed. Most importantly, 2.4 sudden changes in field emission, yielding onset at substantially lower gradient, occur per cavity-century in CEBAF. The phenomenon designated "fratricide" complicates diagnosis but can be dealt with using standard statistical techniques.

Items possibly relevant to future accelerators have been pointed out.  Monitoring of field emission via dedicated x-ray monitors in tunnel is desirable for future accelerators using superconducting RF.

**Acknowledgments**

Programming support for this work has been provided since 2002 by Michele Joyce.  This paper was improved by close review by R. Rimmer, L. Merminga and J. Delayen.

This manuscript has been authored by The Southeastern Universities Research Association, Inc. under Contract No. DE-AC05-84150 with the U.S. Department of Energy. The United States Government retains and the publisher, by accepting the article for publication, acknowledges that the United States Government retains a non-exclusive, paid-up, irrevocable, world wide license to publish or reproduce the published form of this manuscript or allow others to do so, for United States Government purposes.

------------------------------